\documentclass{iau}

\usepackage{amsmath}
\usepackage{graphicx}
\usepackage{multirow}
\usepackage{csquotes}
\usepackage{enumitem}

\begin{document}

\lefttitle{Profitiliotis}
\righttitle{Broadening the Range of SETI Research Ideas}

\jnlPage{1}{7}
\jnlDoiYr{2026}
\doival{10.1017/xxxxx}

\aopheadtitle{Proceedings IAU Symposium}
\editors{J. Haqq-Misra \& R. Kopparapu, eds.}

\title{The Allure of Complete Discovery as Passage: Broadening the Range of SETI Research Ideas via Futures Literacy and Triple-Loop Learning}

\author{George Profitiliotis}
\affiliation{Netherlands Institute of Ecology, Wageningen, Netherlands \& Blue Marble Space, Seattle, Washington, USA}

\begin{abstract}
This paper argues that, although it principally refers to extraterrestrial rather than human affairs, SETI’s imaginary is a social imaginary proper, as it is implicitly linked in an intrinsic, non-trivial, co-constitutive way to a social imaginary of humanity’s future. Specifically, SETI’s imaginary is an imaginary of a society of sapient extraterrestrials that makes possible the achievement of a desirable future of humanity through the former’s discovery by the latter. Moreover, it argues that the range of SETI research ideas, which get bundled in non-fictional conviction narratives promising SETI’s imaginary, is currently limited because actualizing this desirable future state of humanity after such a discovery relies on a “complete discovery”. Finally, an intervention is offered in the form of a hands-on workshop for SETI scientists, which could help them reveal, reframe, and rethink the role that this imaginary for a desirable “post-discovery” future of humanity plays in their present research ideas.
\end{abstract}

\begin{keywords}
SETI, social imaginary, narratives, cosmic mirror, archaeology of the future, technological adolescence, liminality, futures literacy, triple-loop learning
\end{keywords}

\maketitle

\section{Prologue: SETI’s Imaginary, the “Cosmic Mirror”, the “Archaeology of the Future”, Humanity’s “Technological Adolescence”, and the role of a Complete Discovery}

I have previously posited that the overall scientific endeavor of SETI is underpinned by a fundamental collective imaginary that broadly captures the aspired orientation of the SETI research community, thereby anchoring this community, stabilizing its future endeavors, and preventing its fracturing. A reasonable abstraction of the current status of this imaginary is the following: another world beyond Earth where some kind of extraterrestrial intelligent creatures have produced technologies which leave specific traces with attributes that render them detectable by contemporary human observers \citep{proc}. At the same time, many socially-shared non-fictional narratives compete for the conviction of scientists in this community by functioning as alternative future-oriented paths that promise to actionably guide the conduct of SETI towards narrower, more specified, and intelligible “destinations” aligned with the vaguer aspired orientation of that fundamental collective imaginary, namely, more specified kinds of “worlds”, of “extraterrestrial intelligent creatures”, and of “technologies” leaving “traces”.  As I have noted, this broad fundamental imaginary, which is jointly imagined by SETI scholars, concerns extraterrestrial affairs, yet it is also curiously invested with characteristics that SETI scholars desire (or wish) would or would not obtain about our own human civilization too \citep{proc}. This peculiarity calls for further unpacking, which I attempt to advance in this paper. Perhaps counterintuitively, I will argue that, although it principally refers to extraterrestrial rather than human affairs, SETI’s imaginary is a social imaginary proper, as it is implicitly linked in an intrinsic, non-trivial, co-constitutive way to a social imaginary of humanity’s own future. In particular, SETI’s imaginary is an imaginary of a society of sapient extraterrestrials that makes possible the achievement of a desirable future of humanity through the former’s discovery by the latter. Moreover, I will argue that the range of SETI research ideas, which get bundled in the abovementioned non-fictional conviction narratives, is currently constrained and limited because actualizing this desirable future state of humanity after such a discovery relies on a “complete discovery”. I will finally offer a ready-to-deploy educational intervention in the form of a hands-on participatory workshop for SETI scientists, which could help them reveal, reframe, and rethink the role that this imaginary for a desirable “post-discovery” future of humanity plays in their present research ideas.

A social imaginary proper presents a significant “degree of blurring of the boundary between imagining a collective and desiring or wishing it to be real or to have the characteristics one imagines” \citep{sms}. SETI’s imaginary is a social imaginary proper, rather than merely a product of “collaborative imagination”, such as creating a collaborative piece of architecture, or a collective constructed via “socio-cultural imagination”, such as creating a shared folkloric tradition about a fictional fairytale world. SETI’s social imaginary does broadly overlap with these two other kinds of collective imagination, which is not an uncommon case, but also possesses a decisive difference that elevates it to a social imaginary proper \citep{sms}: it appears to somehow involve the phenomena of group belongingness and group identification at the core of its collective imagining. However, the group to which these two phenomena actually refer is not the imagined society of sapient extraterrestrials. These phenomena have been tacitly captured by SETI scholars in what has been historically described as the “Cosmic Mirror” function of SETI: the idea that humans can use the sapient extraterrestrial Other as a vantage point from which to perceive some facets of humanity in comparison to that Other. Although these perceived facets can include not only positive but also negative elements of humanity, such as violence, injustice, misuse, etc., the “Cosmic Mirror” has been described as capable of helping humans realize that they are the same and thus of unifying them \citep{charb}. This alludes to the following: SETI scholars appear to conceptualize humans and sapient extraterrestrials as counter-concepts, i.e., dependent concepts that are relationally articulated and become intelligible by contrast to each other \citep{junge}. This dialectic connection allows SETI scholars to define sapient extraterrestrials through humans and vice versa at the same time. Moreover, SETI has also been historically described as possessing an “Archaeology of the Future” function: it paints a picture of “what we have a chance to become” \citep{morr}. This introduces a temporal asymmetry to this relation. The concept pair of humans and sapient extraterrestrials is already a pair of epistemically asymmetrical counter-concepts, i.e., counter-concepts in which one side—the humans—monopolizes the authority to determine its self-concept and the other’s concept, since the sapient extraterrestrials side is not defined in a self-standing manner, in the absence of any knowledge about them or even about their existence \citep{post}. However, the temporal asymmetry implied in the alleged “Archaeology of the Future” function suggests that the two sides of this pair are not conceptualized as equals either: by presumably being ahead of humans temporally, sapient extraterrestrials are considered more extreme than humans. This imbalance is not fixed in terms of valence: they may be positively or negatively magnified as compared to humans.

The ambivalent asymmetry of the abovementioned pair of counter-concepts reveals an important piece of the puzzle of how the phenomena of group belongingness and group identification underpin SETI’s social imaginary, even though the field at first glance concerns the collective imagining of extraterrestrial affairs. For the SETI scholarly community, the social imaginary of SETI and the social imaginary of humanity’s future operate like communicating vessels implicitly linked in a co-constitutive way through that core pair of asymmetrical counter-concepts. In particular, the explicit collective imagining of a society of sapient extraterrestrials appears to function sometimes as an encouraging example and other times as a deterrent example for an implicit collective imagining of humanity’s future, while at the same time existing knowledge of humanity’s past and present implicitly provides building blocks for generating those explicit imaginings of such a society of sapient extraterrestrials. It is through this co-constitutive function that the phenomena of group belongingness and group identification play their decisive role in SETI’s social imaginary at an implicit level: SETI scholars appear to identify with the social group of humanity as a whole, which provides them with a sense of belongingness to this group. Because of that co-constitutive relationship, the effects of these two phenomena also seep into SETI scholars’ collective imagining processes about sapient extraterrestrials and give rise to SETI’s social imaginary proper, even though they actually originate in a deeper, coupled social imaginary of humanity’s own future. I posit that these two phenomena also underlie a particular shared motivating goal—or motive—of SETI scholars for the group of humanity, which can be explicated through SETI scholars’ idea of humanity’s “Technological Adolescence”: the view that our contemporary epoch is a “dangerous moment” for humanity because humanity will either manage “to avoid self-destruction”, and thus “survive” it, or not \citep{sag}. Indeed, the existence of such a danger of annihilation has also been captured in the conjecture of a “Great Filter” that may have been part of humanity’s past or may still be lurking in its future \citep{hmks}. I posit that overcoming this “Technological Adolescence” via evading this danger is the highest-order, implicit, shared motivating goal of SETI scholars for the group of humanity, i.e., either a common desire or a common wish, depending on whether its attainment is correspondingly viewed as more possible or less possible \citep{mica}. 

The uncertainty surrounding the potential outcomes of this “Technological Adolescence” epoch, namely, the uncertainty about whether humanity will avoid that self-destruction (which favors the motivating goal) or not (which opposes the motivating goal), also gives rise to the anticipatory emotion of anxiety within the community of SETI scholars, i.e., an alternation or mixture of the conflicting anticipatory emotions of hope and fear. At the same time, because of this uncertainty, two other, instrumental goals get subordinated to that motivating goal \citep{mica}: the goal to find out whether the aforementioned danger will occur and whether humanity can cope with it, and the goal that this danger does not occur. Curiously, for SETI scholars, these two goals become further hierarchically linked to one lower-order, subordinated, instrumental goal: discovering sapient extraterrestrials sooner rather than later via SETI. Thanks to a peculiar cognitive tendency that will be unpacked in the next paragraph, these additional lower-order levels of that goal hierarchy implicitly echo the motivational force of the high-order motivating goal among SETI scholars, impelling them to act by means of conducting SETI. This motivational feedback influence from the desirable “post-discovery” future of a post-adolescent humanity to the “pre-discovery” present of conducting SETI allows an expansion of the anticipatory emotion of anxiety from its original locus, i.e., humanity’s future, to the extraterrestrial locus. Specifically, the anxiety emotion experienced in the present is initially elicited implicitly by the mental representation of the possible future outcomes of humanity’s current adolescence. Afterwards, however, because of a lack of an explicit awareness of its initial source, it accomplishes a premonitory function, though this time vis-à-vis sapient extraterrestrials: it elicits some fitting mental representations of them which can act as implicit “mini-theories” that offer satisfying yet incorrect explanations for why this uncertainty-infused blend of fear and hope is experienced within the SETI community \citep{casmic2011}. These “mini-theories” that are allowed by this expansion of the anxiety emotion are an imaginational feedforward influence comprising the generation of explicit imaginings of sapient extraterrestrials through selected building blocks offered by existing knowledge of humanity’s past and present, which then function as encouraging or deterrent examples for implicit collective imaginings of humanity’s future. In other words, this expansion of the anxiety emotion amplifies the implicit bidirectional interplay between the mental representations of sapient extraterrestrials to be discovered and of humanity’s post-adolescent future to be achieved.

To rephrase and elaborate on this, the SETI community’s failure to explicitly acknowledge the mentally-represented uncertain future of humanity as the actual origin of the anxiety emotion that is anticipatorily experienced in the present allows this emotion to be implicitly incorrectly explained—due to its premonitory function—as being caused by some fittingly-elicited mental representations of sapient extraterrestrials waiting to be discovered, namely representations that are suitably constructed to be capable of generating such an uncertainty-infused blend of fear and hope through selected building blocks derived from existing knowledge of humanity’s past and present. In turn, those mental representations of sapient extraterrestrials warp the mental representations of humanity’s future through the pre-existing co-constitutive link between the concept pair of humans and sapient extraterrestrials as asymmetrical counter-concepts, as explained previously. However, this emotion-based amplification of that conceptual connection between the social imaginary of SETI and the social imaginary of humanity’s future, which is related to the motivational feedback influence and to the imaginational feedforward influence, harbors errors. On the one hand, the collective imagining of sapient extraterrestrials by SETI scholars is plagued by the information processing bias of selective attention, which is tacitly constricting scholars’ internally-directed attention towards some specific, conveniently understood ideas drawn from humanity’s past and present, due to the presence of that deeper motivating goal for obtaining their lower-order instrumental goal of discovering via SETI sapient extraterrestrials sooner rather than later. This is because attention is generally known to be susceptible to getting preconsciously steered by chronic and even temporary motivating goals \citep{mosk}, while internally-directed attention specifically, i.e., a focus on self-generated mental representations, plays a key role in supporting imagination and creative cognition \citep{benedek}. This is augmented by the emotion-related need for constructing suitable mental representations of sapient extraterrestrials to be capable of erroneously generating the uncertainty-infused blend of fear and hope and thus of explaining why it is being experienced. On the other hand, the goal architecture generated by SETI scholars for humanity as a group is plagued by the cognitive distortion of conditional goal setting, which is tacitly constricting scholars’ perception of the achievement of the lower-order instrumental goal of discovering via SETI sapient extraterrestrials sooner rather than later as necessary for obtaining the top-level end goal of overcoming humanity’s adolescent period. This is because conditional goal setting, i.e., the tendency to regard high-order abstract goals as dependent on the achievement of lower-order concrete goals, has been shown to occur more strongly in times of increased hopelessness about the future, namely, when people no longer see their high-order goals as attainable despite still remaining committed to them \citep{hama}, while this conditional link between the high- and low-order goals increases resistance to goal reorganization and exaggerates the significance of overcoming obstacles to the attainment of the low-order goals as if directly protecting the high-order goals \citep{crane}. This augments the outcome of the uncertainty-infused blend of fear and hope emotions, generated initially by the mentally-represented uncertain future of humanity, through the inflated rendering of any hurdles in discovering sapient extraterrestrials via SETI as directly fueling the perpetuation of doubts and questions over humanity’s own future survival. 

Overall, the anticipatory anxiety emotion, the selective attention bias, and the conditional goal setting distortion that originate in the interplay between the social imaginary of SETI and the social imaginary of humanity’s future do not stay confined at the more abstract level of the imaginary but spread into the more actionable level of the socially-shared non-fictional narratives that offer alternative paths forward for the conduct of SETI. This is because the primary purpose of any future-oriented non-fictional narrative is to connect the present situation with a corresponding imaginary. All such narratives inherently function by steering people’s attention to a selected subset of ideas that appear to be aligned with a specific imaginary and can be causally combined into various possible paths towards achieving it, one of which is usually spotlighted through emotional commitment as the most actionable course that can lead communities to achieving  said imaginary, if collectively pursued \citep{proc}. In other words, such narratives are fundamentally vulnerable to errors related to attention, goal setting, and emotional commitment. Thus, non-fictional narratives about the conduct of SETI essentially channel the anticipatory anxiety emotion, the selective attention bias, and the conditional goal setting distortion from the abstract level of the imaginary to the actionable level of how to conduct SETI in order to realize SETI’s imaginary. This constrains the various specified kinds of “worlds”, of “extraterrestrial intelligent creatures”, and of “technologies” leaving “traces” that each narrative champions as its preferred “destination” that can fulfill the promise of reaching the vaguer fundamental collective imaginary of SETI. Although at the explicit level they address matters related to their championed path of conducting SETI, at the implicit level these narratives fulfill their promise by delivering a “destination” that does not embody a “pure” imaginary of SETI but rather an imaginary of SETI that is intrinsically linked to an imaginary of humanity’s future, in the ways described in the previous paragraphs.

Before concluding this prologue, there is one last point that needs unpacking. As discussed above, the epistemic goal to find out whether the danger of annihilation related to humanity’s “Technological Adolescence” will occur and whether humanity can cope with it, and the pragmatic goal that this danger does not occur become further hierarchically dependent on the instrumental goal of discovering via SETI sapient extraterrestrials sooner rather than later. In turn, this impels SETI scholars to act by means of conducting SETI. In other words, the conduct of SETI appears to be viewed by SETI scholars as a way to pursue both that epistemic and that pragmatic goal at the same time. The transition from the so-called “Technological Adolescent Age” to the “Technological Mature Age” of living in harmony with one’s species and natural environment has been described as a possibly typical procedure for all technology-capable civilizations \citep{lema}. This means that a successful SETI discovery is thought to be able to contribute to achieving the epistemic goal of accomplishing some understanding of the risk of annihilation through some comparative study between the human society and the discovered extraterrestrial one. In parallel, a successful SETI discovery has also been described as a catalyst for forcibly changing humanity into acting with more compassion and humility \citep{denn}. This means that such a discovery is at the same time thought to be able to somehow contribute to also achieving the pragmatic goal of exiting this period of adolescence successfully without being thwarted by annihilation. Although the presumed “how” behind the latter part has not been clearly articulated, I argue that it is possible to explicate its implicitly accepted mechanism by referring again to the ideas of the “Cosmic Mirror”, which can unify humanity, and the “Archaeology of the Future”, which presents humanity with what it has a chance to become. In particular, I posit that the implicit mechanism which appears to connect these ideas with humanity’s maturation is that the conduct of SETI is thought to play a key role within a peculiar “rite of passage” for the whole of humanity. A rite of passage is a tripartite ritual scheme of separation, transition, and incorporation undertaken by individuals or entire communities when passing from “one situation to another or from one cosmic or social world to another” \citep{vang}. The intermediate period of transition in this tripartite scheme has been termed “liminality”, i.e., a threshold situation of “interstructural” suspension “betwixt and between” two stable states \citep{tur}. Although originally studied in small-scale, traditional communities, this idea of passage has been theorized to also be applicable to entire contemporary societies. When an entire society enters a liminal situation, it “needs orientation”, which is sought through “imitative processes” that resort to looking to “‘others’ for guidance”. If this liminal situation does not get resolved and a new stability is not found, liminality across society can become lasting \citep{szab}. I posit that the SETI community’s ideas of “Technological Adolescence” and “Technological Mature Age” correspondingly reflect the “transition” state and the “incorporation” state for humanity as a whole, while the conduct of SETI is implicitly viewed as able to offer the necessary orientation for resolving humanity’s lasting liminality (adolescence) by resorting to ostensible sapient extraterrestrials for deriving a sort of imitative guidance. Of course, this implicit idea also requires the tacit perception of humanity’s temporal endurance as an overall group, a perception which has been shown to be enhanced in situations of significant existential anxiety \citep{sani}.

For the reasons outlined above, I suggest that the predominant implicit focus of socially-shared non-fictional narratives competing for the conviction of SETI scientists \citep{proc} in how to conduct SETI is the production of “complete discoveries”, namely, discoveries that can swiftly pass through all the three stages of “detection”, “interpretation”, and “basic understanding” \citep{dicka}. This is because complete discoveries are tacitly seen as able to resolve humanity’s passage through adolescence by providing knowledge about a society of sapient extraterrestrials, which can then be used as an encouraging or deterrent example from which to derive a model that can be concretely followed by humanity to exit its liminal situation in a transformed state \citep{szaa}. Therefore, non-fictional narratives about the conduct of SETI favor research ideas about searching for phenomena that can sooner rather than later fulfill all the following criteria: their detection will not be one-off; their interpretation as being artificial will be clearly unambiguous; and the basic understanding of the causal mechanism behind them will be eventually theoretically unpacked by conventional science \citep{dicka}. These criteria appear to reverberate across discussions about what kinds of search proposals have “merit” \citep{shei}. In other words, the socially-shared non-fictional narratives competing for the conviction of SETI scientists by functioning as alternative paths to a complete discovery of a society of sapient extraterrestrials, sooner rather than later, are constraining the conduct of SETI by disfavoring research ideas about searching for fundamentally strange phenomena whose discovery could remain incomplete for a protracted period of time. Lastly, it should be noted that this skewing imposed on the range of possible research ideas by the acceptable narratives in which they can get bundled is not fixed and static. When left to their own devices, social imaginaries can gradually spiral due to group polarization towards more intensified collective expressions of desired or wished-for elements. Not only can these collective imaginings then take root in individuals’ minds, to the point of becoming indistinguishable from individual imaginings, but also their intensified desired or wished-for elements can become hard to discern from their factual ones \citep{sms}. Such an intensification of collective expressions of desired or wished-for elements in SETI’s imaginary, due to the linked imaginary of humanity’s future, can propagate across the narratives about the conduct of SETI, siphoned through each narrative’s preferred “destination” which is a more specific translation of SETI’s broader imaginary. In turn, this means that the range of SETI research ideas that can get bundled into the at the time acceptable narratives can accordingly bend towards less diversity of imaginings and less discernibility of factuality from desirability and wishfulness.

\section{Broadening the Range of SETI Research Ideas: A Deployable Futures Literacy and Triple-Loop Learning Workshop Workflow}

One way to counteract this detrimental bending that can limit the conduct of SETI is to help SETI scholars reveal, reframe, and rethink the role that the aforementioned imaginary for a desirable “post-discovery” future of humanity plays in their present research ideas. To this end, this section presents a ready-to-deploy intervention in the form of a hands-on, participatory, educational workshop workflow drawing on triple-loop learning and futures literacy. Triple-loop learning is a form of learning that recursively operates on top of double- and single-loop learning. Single-loop learning is a form of learning that focuses on acting upon a problem at hand by changing the methods and tactics used to address it, drawing from a fixed set of alternative options. Double-loop learning goes one level of complexity deeper and focuses on questioning and challenging the fundamental assumptions and logics that underlie one’s understanding of a problem at hand, thereby facilitating the change of the entire set of alternative options that was regarded as fixed. Triple-loop learning goes another level of complexity deeper and focuses on continuously employing active awareness to discern and purposely reshape one’s motives that influence one’s ways of understanding of and acting upon a problem at hand, thereby facilitating the change of the whole apparatus through which entire sets of alternative options can come to be \citep{kwni}. The proposed educational workshop workflow operationalizes triple-loop learning by means of a series of activities that follow the general structure of a “Futures Literacy Laboratory-Novelty” (FLL-N). 

The FLL-N is a tailorable, hands-on, interactive tool that has been specifically developed to promote in various contexts the enhancement of futures literacy among its participants, i.e., the capability to bring the future into the present through imagination—in other words, to “use-the-future”—for different reasons and in a variety of ways \citep{mila}. The FLL-N’s tailorable general structure comprises three consecutive phases. Firstly, the “Reveal” phase helps participants make explicit their currently-held implicit ideas about the probable and desirable futures and then interrogate the fundamental assumptions that sustain them. Then, the “Reframe” phase exposes participants to an unfamiliar future that is completely freed from ideas about what is considered probable and what is considered desired/wished-for and helps them rehearse and grasp a bewildering world. Finally, the “Rethink” phase prompts participants to compare and contrast their experiences in the previous two phases, to reflect on new facets of the future that were illuminated by this process, and to realize that specific ideas about what is considered probable and what is considered desired/wished-for tend to tacitly constrain the way the present is being seen \citep{milb}. Previously-published FLL-N-type workshop workflows for the SETI scholars community  focused on enhancing their preparedness towards the short-term repercussions immediately after a SETI-relevant detection and its communication \citep{proa} and on augmenting their moral imagination towards pre-empting the monsterization of extraterrestrial lifeforms during the critical first moments of a potential straightforward encounter \citep{prob}. The workflow of the “Broadening the Range of SETI Research Ideas” workshop tailors the general tripartite structure of the FLL-N to the particular context of helping SETI scholars relax the limitations that are posed to their present research ideas for the conduct of SETI due to a desirable “post-discovery” future of humanity made possible by the long-term repercussions of a SETI-relevant “complete discovery”. This workshop is designed to accommodate individual and plenary work for 50 participants with a single facilitator, to keep the overall duration limited to 2 hours. For the workshop to be interactively implemented in virtual, physical, and hybrid formats, it is important that all the individual free-form text writing and voting outcomes are presented in real time to the plenary, so that participants can see others’ votes and read their thoughts as the process unfolds. The workflow is presented below, while its envisioned effects and their mechanics are briefly discussed in the epilogue section.

The content of the “Reveal” phase of the “Broadening the Range of SETI Research Ideas” workshop is structured as follows. Firstly, participants are provided with a drawing of an Overton Window \citep{french} marked on an one-dimensional axis going from “Fully data-driven SETI research (empirico-inductive)” to “Fully hypothesis-driven SETI research (hypothetico-deductive)” \citep{garo}. This axis is qualitatively segmented into 11 levels: the middle level is marked as “NORM”, which gradually grows on both sides towards the abovementioned two extremes following the procession “POPULAR”, “SENSIBLE”, “ACCEPTABLE”, “RADICAL”, and “UNTHINKABLE”. The Overton Window is visibly marked as comprising the “ACCEPTABLE”, “SENSIBLE”, “POPULAR”, “NORM”, “POPULAR”, “SENSIBLE”, “ACCEPTABLE” parts of this axis. Participants are then given the following instructions to work on this axis via free-form text.
\vspace{-2ex}
\begin{quote}
We are now in 2026. This is the present day. The Search for Extraterrestrial Technology-capable Intelligence, or for Technosignatures, involves a portfolio of ideas about what to search for, how to search for it, why, etc. As with other scientific fields, these ideas coalesce around a spectrum ranging from fully hypothesis-driven research to fully data-driven research. However, people in the SETI academic community have raised concerns about a possible “failure of imagination” \citep{dickb} that could be limiting our ideas. In this activity, we will silently make an initial attempt to capture the landscape of SETI research ideas that have been circulating around the SETI academic community, including the “whats”, the “hows”, the “whys”, or any other pertinent aspects, using an Overton Window as a guide. The Overton Window is a conceptual model that captures the range of ideas for a particular subject which are present in a community’s discourse at a given time and arranges them along a spectrum of acceptability. According to this model, the strategic viability and success of an idea depend on the particular climate of mainstream community acceptability, i.e., its position with respect to the Window, at a given time. The SETI academic community has historically felt the detriments and benefits of such effects as expressed through the scope of various funding institutions’ calls for research proposals and/or any subsequent proposal evaluations, among others. Drawing on your own knowledge of and experience with the SETI research landscape, silently think about and write down the various circulating SETI research ideas by orienting them along the two poles of the following Overton Window: from ideas leaning towards fully data-driven SETI research to ideas leaning towards fully hypothesis-driven SETI research. When deciding an idea’s level of acceptability, express your own judgment rather than your guess about an ostensible overall judgment of the community! There is no prescribed order for working on the levels between the two poles: you can start from whichever level you want and jump to any other level at any time! Importantly, do not censor yourselves! We also need radical and unthinkable ideas to reflect on later! Feel free to also read others’ ideas for cross-pollination!
\end{quote}
\vspace{-2ex}
After the participants have written down their ideas on the axis, they are given the following instructions to continue working through the “Reveal” phase of the workshop by traveling to a desirable future. 
\vspace{-2ex}
\begin{quote}
Let’s travel to the future! Please close your eyes and try to imagine the following. You are now immediately transported several years into the future. In this future world, SETI research has finally succeeded, thanks to an initial putative detection of a particular phenomenon in the year 2026, and the dust of its success has settled. The whole chain of scientific follow-ups and non-scientific implications stemming from that initial putative detection and extending all the way to this future world in which you find yourself \textit{went exactly as you wanted it to go}. Please picture yourself exploring this future’s versions of various locations from your everyday activities, such as your neighborhood, your workplace, your favorite community place, etc. How does people’s daily life look like in \textit{your most desirable version of the future} after the dust of SETI’s success has settled? Write a short note to yourself about something that piqued your interest in this world! It can be anything you saw, or heard, or otherwise experienced, which you thought was noteworthy! Start by what year it is! Have a look around and read others’ notes from their own individual futures!
\end{quote}
\vspace{-2ex}
After the participants have written their short notes, they are given the following instructions. 
\vspace{-2ex}
\begin{quote}
You are still in \textit{your individual most desirable version of the future} after the dust of SETI’s success has settled. Remember: this success was owed to an initial putative detection of a particular phenomenon in the year 2026. As you explore this future world, you come across an official infographic published by a reputable SETI research organization which describes the complete chronicle of this success. Based on what was highlighted in that infographic, write a short note to yourself describing (a) what was the particular phenomenon that was putatively first detected in the year 2026, and (b) how exactly did that lead to SETI’s success!
\end{quote}
\vspace{-2ex}
After the participants have written their additional short notes, they are given the following instructions. 
\vspace{-2ex}
\begin{quote}
Now that you explored it, reflect on \textit{your individual most desirable version of the future} after the dust of SETI’s success has settled. Which of these generic future archetypes best describes this future of yours? Place one anonymous vote to one archetype of your choice!
\end{quote}
\vspace{-2ex}
They are then provided with the following descriptions of Dator’s four archetypal images of the future \citep{dator} to be used in their voting:
\vspace{-2ex}
\begin{quote}
\begin{itemize}[leftmargin=*]
    \item {Decline archetype (↓): Relative to March 2026, this future after the dust of SETI’s success has settled features humanity in a state of decline, or even collapse}
    \item Growth archetype (↑): Relative to March 2026, this future after the dust of SETI’s success has settled features humanity in a state of continuous “business-as-usual” growth
    \item Equilibrium archetype ($\sim$): Relative to March 2026, this future after the dust of SETI’s success has settled features humanity in a state of limitation, balance, maintenance, or control
    \item Transformation archetype (!): Relative to March 2026, this future after the dust of SETI’s success has settled features humanity in a novel, radically different state
\end{itemize}
\end{quote}
\vspace{-2ex}
After everyone has voted on the archetypes, participants are given the following instructions. 
\vspace{-2ex}
\begin{quote}
What was the deeper significance of SETI’s success in this most desirable version of your future? Remember: the whole chain of scientific follow-ups and non-scientific implications stemming from that initial putative detection and extending all the way to this future \textit{went exactly as you wanted it to go}. Place one anonymous dot on the appropriate point framed by the two axes below!
\end{quote}
\vspace{-2ex}
They are then provided with two intersecting, perpendicular axes to be qualitatively used as a Cartesian coordinate system. The vertical axis goes from “SETI’s success meant that \textit{there is a} universal course preordaining the development of all technospheres, terrestrial and extraterrestrial alike” to “SETI’s success meant that \textit{there is no} universal course preordaining the development of all technospheres, terrestrial and extraterrestrial alike”. The horizontal axis goes from “SETI’s success meant that the source of the creative processes behind an extraterrestrial technology had been \textit{concentrated within one or more sapient extraterrestrial entities of a single kind which independently pursued their distinctive} interests” to “SETI’s success meant that the source of the creative processes behind an extraterrestrial technology had been \textit{diffused across a web of one or more extraterrestrial entities of different kinds, sapient and non-sapient alike, which jointly pursued their varying} interests”. After everyone has placed their dot on the two perpendicular axes, participants are given the following instructions. 
\vspace{-2ex}
\begin{quote}
Finally, reflect on the articulation between the initial putative detection in the year 2026 and SETI’s success in this \textit{most desirable version of your future}, based on the infographic you came across. Remember: the whole chain of scientific follow-ups and non-scientific implications stemming from that initial putative detection and extending all the way to this future \textit{went exactly as you wanted it to go}.
\end{quote}
\vspace{-2ex}
They are then provided with the following incomplete timeline:
(March 2026) → (Initial putative detection in 2026)—( ? )—(SETI’s success) → (The \textit{desirable} future to which you travelled). Based on this, participants are then asked to complete the missing part of the timeline denoted by the question mark between the milestones “Initial putative detection in 2026” and “SETI’s success”, by selecting the option they perceive as the most fitting out of a list of 14 items, as explained in the following instructions.
\vspace{-2ex}
\begin{quote}
Which of the following generic timeline descriptions best describes this articulation? Place one anonymous vote on the segment description of your choice (the one which, according to your note on the infographic you came across, covers the missing bit of the timeline)!
\end{quote}
\vspace{-2ex}
They are then provided with this list of generic timeline descriptions of a discovery's conceivable course.
\vspace{-2ex}
\begin{quote}
\begin{enumerate}
    \item The putatively detected phenomenon was immediately confirmed → It was instantly obvious that it was a technosignature → The mechanism that produced this TS was explainable straight away.
    \item The putatively detected phenomenon was immediately confirmed → It was instantly obvious that it was a technosignature → Then, it took some more time to figure out the best explanation behind the mechanism that produced this TS.
    \item The putatively detected phenomenon was immediately confirmed → It was instantly obvious that it was a technosignature → Then, it was impossible to figure out the best explanation behind the mechanism that produced this TS.
    \item The putatively detected phenomenon was immediately confirmed → Afterwards, it took some time to conclude that it would best be interpreted as being a technosignature → The mechanism that produced this TS was explainable straight away.
    \item The putatively detected phenomenon was immediately confirmed → Afterwards, it took some time to conclude that it would best be interpreted as being a technosignature → Then, it took some more time to figure out the best explanation behind the mechanism that produced this TS.
    \item The putatively detected phenomenon was immediately confirmed → Afterwards, it took some time to conclude that it would best be interpreted as being a technosignature → Then, it was impossible to figure out the best explanation behind the mechanism that produced this TS.
    \item The putatively detected phenomenon was immediately confirmed → Afterwards, it was impossible to conclude that it would best be interpreted as being a technosignature.
    \item It took some time to confirm the putatively detected phenomenon → It was instantly obvious that it was a technosignature → The mechanism that produced this TS was explainable straight away.
    \item It took some time to confirm the putatively detected phenomenon → It was instantly obvious that it was a technosignature → Then, it took some more time to figure out the best explanation behind the mechanism that produced this TS.
    \item It took some time to confirm the putatively detected phenomenon → It was instantly obvious that it was a technosignature → Then, it was impossible to figure out the best explanation behind the mechanism that produced this TS.
    \item It took some time to confirm the putatively detected phenomenon → Afterwards, it took some time to conclude that it would best be interpreted as being a technosignature → The mechanism that produced this TS was explainable straight away.
    \item It took some time to confirm the putatively detected phenomenon → Afterwards, it took some time to conclude that it would best be interpreted as being a technosignature → Then, it took some more time to figure out the best explanation behind the mechanism that produced this TS.
    \item It took some time to confirm the putatively detected phenomenon → Afterwards, it took some time to conclude that it would best be interpreted as being a technosignature → Then, it was impossible to figure out the best explanation behind the mechanism that produced this TS.
    \item It took some time to confirm the putatively detected phenomenon → Afterwards, it was impossible to conclude that it would best be interpreted as being a technosignature.
\end{enumerate}
\end{quote}
\vspace{-2ex}
After everyone has voted on the missing bit of the timeline, participants are given the following instructions.
\vspace{-2ex}
\begin{quote}
Silently contemplate your personal choices in the previous three steps through the lens of this question: What are the overall foundations that hold together (a) your timeline description from the initial putative detection to SETI’s success, (b) your deeper significance of SETI’s success, and (c) your archetype of that future after the dust of SETI’s success had settled? Choose any that apply from the following “List of Foundations”. You may select more than one.
\end{quote}
\vspace{-2ex}
They are then provided with this “List of Foundations” \citep{prob}:
\vspace{-2ex}
\begin{quote}
\begin{itemize}[leftmargin=*]
    \item I built on forecasts, predictions, or projections that extrapolate from the past to the future
    \item I built on ideological or religious accounts that predetermine a fixed future destiny
    \item I speculated on future creative interventions to be undertaken by an individual or group to solve some currently known issues in their external environment
    \item I speculated on future self-improvements to be undertaken by an individual or group to overcome some currently known shortcomings within themselves
    \item I relied on a systematic consideration of the possible interplays of various factors to notice novel but recurring patterns as they were emerging
    \item I relied on active, alert awareness and mindfulness to notice something rare or unique as it was emerging
\end{itemize}
\end{quote}
\vspace{-2ex}
After everyone has voted on the “List of Foundations”, participants are given the following instructions.
\vspace{-2ex}
\begin{quote}
We now leave this \textit{desirable} future behind temporarily, as we continue our journey into the futures!
\end{quote}
\vspace{-2ex}
At this point, the exact sequence of activities of the desirable future that was described above is repeated once more, but this time for the probable future. To this end, participants are given the same instructions as before, but this time the phrases “most desirable version of the future” and “desirable future” are replaced by “most probable version of the future” and “probable future”, while the phrase “went exactly as you wanted it to go” is replaced by “went exactly as you believed it would go”. These changes across the instructions make possible the elicitation of insights for the probable future that are directly comparable with those for the desirable future.

With the completion of the probable future round, the workshop transitions from the “Reveal” phase to the “Reframe” phase. This second phase requires participants to be transferred to a strange future of 2030: the Future of the Apparition. To this end, participants are given the following instructions.
\vspace{-2ex}
\begin{displayquote}\setlength{\parindent}{1em}
We now leave this probable future behind temporarily, as we continue our journey into the futures! Let’s travel to the future one last time! Please close your eyes and try to imagine the following. You are now immediately transported into the year 2030. In this future world, SETI research has definitively delivered a confirmed detection of a particular phenomenon, which was putatively first detected in 2026 and got independently corroborated in the same year. However, the scientific follow-ups and non-scientific implications stemming from that confirmed detection went neither as you believed they would go nor as you wanted them to go. You find yourself in a strange future of 2030: a future in flux. This is what happened before you arrived here.

In early 2026, amateur astronomers analyzing publicly available coronagraph images from the Solar and Heliospheric Observatory satellite (SOHO) identified a relatively small-sized, very low inclination object without a visible coma or tail. It was approaching an extremely small perihelion distance of less than 2 solar radii from the Sun’s surface, based on its preliminary orbital parameters that were calculated from the short observational arc available. This was consistent with the behavior of some rare cases of sungrazing comets, although the majority of usual sungrazers tend to have much higher inclinations.

Unlike most sungrazers, the object did not disintegrate: it re-emerged in post-perihelion frames, brighter than before. Astrometry from ground-based observatories filled in its arc sufficiently to show that the initial, crudely-calculated, slightly parabolic trajectory had now changed into a slightly elliptic one, indicating heliocentric velocity reduction through non-gravitational forces at perihelion. Calculations showed that the object’s new orbit would intersect Earth’s orbital distance near 1 AU some months later.

Some weeks after its re-emergence from perihelion, the object exhibited another unexpected non-gravitational change of velocity that affected its trajectory, creating a significant probability of a forthcoming gravitationally-important interaction with the Earth-Moon system at the time of its predicted crossing of the ecliptic. A single broadband radio burst coincided with this change. This was initially detected by a ground-based observatory and later corroborated through a retrospective analysis of data from other radio facilities that were monitoring the same sector.

Then came the Apparition. Just days before its projected closest approach to the Earth-Moon system, and its highly probable ballistic capture by it due to the favorable timing and relative velocity, the object exhibited a sudden and very strong illumination in visible light.

Some years later, i.e., in this strange future of 2030, the Apparition is a new, peculiarly illuminating companion in our sky. It appears to have, at least temporarily, entered a very-high-altitude elliptical orbit around the Earth, farther than a geosynchronous orbit, after a chaotic three-body capture which endowed it with a medium inclination that oscillates due to lunar perturbations. 

This orbit, combined with Earth’s rotation, makes the Apparition visible from all places on Earth at various times during its orbital period.

Here is a recent picture of the Apparition taken via a mobile phone in Athens, Greece, in this strange future.
\end{displayquote}
\vspace{-4ex}
Participants are then provided with the fictitious picture of the Apparition, available within the workshop’s openly available companion image album: https://doi.org/10.5281/zenodo.20594903 . They are then given the following instructions. 
\vspace{+4ex}
\begin{quote}
Despite its definitively confirmed detection and the many years that have passed since then, the Apparition has not yet been definitively interpreted as either an extraterrestrial engineered phenomenon or not, neither has it been definitively understood scientifically. As visitors in this strange future, you encounter strange everyday things and strange everyday places. Here are some glimpses of places and things you recently encountered. All these places and things are somehow related to the consequences of the Apparition. Keep in mind these images. Imagine you are experiencing the people, the systems, the places, the things, and the events of this strange future world of the Apparition. What is going on there? How does people’s daily life look like in this world? Write a short note to yourself about something that piqued your interest in this world! It can be anything you saw, or heard, or otherwise experienced, which you thought was noteworthy! Feel free to have a look around and read others’ notes from this strange future which hosts all of you! 
\end{quote}
\vspace{-2ex}
Participants are then provided with the nine fictitious context pictures of this strange future world, also available within the workshop’s openly available companion image album: https://doi.org/10.5281/zenodo.20594903 . 
After the participants have written their short notes, they are given the following instructions. 
\vspace{-2ex}
\begin{quote}
This Future of the Apparition which you explored above has become an extended borderline period for human institutions, full of disruption, instability, and ambivalence. Because of its inability to deliver a definitive interpretation of the Apparition as either an extraterrestrial engineered phenomenon or not, let alone to advance its scientific understanding, the field of SETI research is also at a crossroads. Keep in mind this strange world, as we continue below. As scientists in this strange future of 2030, you are now participating in a global high-level forum of SETI researchers, the first one after the Apparition. This forum will produce a manifesto on the “Future of SETI Science” and is live-streamed all over the world. As authenticity in science is highly valued in this world, each forum activity is a surprise both for you and for the remote audience. For the first unexpected activity of this forum, you are asked to anonymously provide honest opinions about what the Future of SETI Science must and must not involve from the specific viewpoint of any of the colored six “thinking hats” (by Edward de Bono) that you can find below! Write your honest opinions under any of the six colored hats and also read those of others!
\end{quote}
\vspace{-2ex}
They are then provided with the following descriptions of de Bono’s six thinking hats \citep{debono} to be used when expressing their opinions:
\vspace{-2ex}
\begin{quote}
\begin{itemize}
    \item Blue hat: focus on planning, processes, meta-reasoning
    \item White hat: focus on facts, data, objective information
    \item Red hat: focus on feelings, hunches, gut instinct  
    \item Green hat: focus on ideas, possibilities, creative alternatives  
    \item Yellow hat: focus on logical reasons for usefulness, benefits, positive points 
    \item Black hat: focus on logical reasons for cautions, difficulties, dangers
\end{itemize}
\end{quote}
\vspace{-2ex}
After everyone has written their honest opinions under the appropriate thinking hats, participants are given the following instructions.
\vspace{-2ex}
\begin{quote}
For the second unexpected activity of this “Future of SETI Science” forum, you are asked to leave your colored “thinking hats” behind, although you can use anything you learned by doing the previous activity to your advantage. Now, you are given a unique opportunity to improvise a 30-second nano-pitch for a SETI research proposal to secure micro-donations from the global audience! Your pitch will be automatically and accurately translated by AI, but it must be social-media-ready and must consist of three convincing phrases: your SETI research proposal, its future impact within academia, and its future impact beyond academia. There will be no subsequent assessment stage after that pitch: you will be evaluated by the crowdfunders in real time!
\end{quote}
\vspace{-2ex}
They are then provided with a form containing three focal points to be filled in with appropriately convincing phrases: “Your SETI Research Proposal”; “Its Prospective Knowledge Impact (impact within academia)”; and “Its Prospective Real-World Impact (impact beyond academia)”. After everyone has written their nano-pitch in their form by filling in the three focal points, participants are given the following instructions.
\vspace{-2ex}
\begin{quote}
Silently contemplate the thought process which helped you compose the content of your nano-pitch during the previous activity through the lens of this question: What are the overall foundations that hold together (a) the scientific path leading to your proposal’s future success, (b) the future knowledge impact of your proposal’s future success, and (c) the future real-world impact of your proposal’s future success? Choose any that apply from the following “List of Foundations”. You may select more than one.
\end{quote}
\vspace{-2ex}
They are then provided with the “List of Foundations” as before. With the completion of the strange future round, the workshop transitions from the “Reframe” phase to the “Rethink” phase. This third and final phase requires participants to scavenge across the three kinds of futures they visited during the previous activities. To this end, participants are given the following instructions.
\vspace{-2ex}
\begin{quote}
We now leave this future behind, as we continue our journey back to the present! Your journey into the future(s) has almost reached its end. Reflect on all the previous activities by comparing and contrasting the desirable, the probable, and the strange futures. What new insights have you discovered while accessing that strange future of the Apparition, as compared to what you initially considered probable and/or desirable in terms of scientific follow-ups and non-scientific implications stemming from an initial putative SETI-relevant detection? These insights can be related both to how and to what we anticipate when we engage with the future!
\end{quote}
\vspace{-2ex}
After everyone has written their insights, participants are given the following instructions.
\vspace{-2ex}
\begin{quote}
You are now back to the present day! Look again at the Overton Window for the SETI research landscape in 2026, as captured in your own words at the beginning of this workshop. Do any of the collective insights that you scavenged from your trips to the futures help you see any present matters related to SETI research in a new light? Do you sense any “filters” imposed by desirable and/or probable futures on the form of our Overton Window, thereby unwittingly modulating the salience of our ideas? What actions could be taken right now, in the present, to bypass those “filters” and fundamentally alter our Overton Window to consciously broaden what the present affords to the Search?
\end{quote}
\vspace{-2ex}
With the completion of the participants’ collective rethinking of the Overton Window for the SETI research landscape and their overall reflection on the “filters” imposed on it by the future(s), the “Rethink” phase concludes, thereby bringing the workshop to a close.

\section{Epilogue: Consciously Alternating Between the Two Subtypes of Deliberate Sensemaking to Benefit SETI Research}

The envisioned effects of the workshop that was presented in the previous section are twofold. On the one hand, single- and double-loop learning effects are envisioned to occur via the mechanic of co-creating the Overton Window and revising it after the series of mental time-travels to the different kinds of futures. On the other hand, triple-loop learning effects are envisioned to occur on top of these two loops via the mechanics of dissecting the anticipatory assumption clusters behind those mental time-travels to different kinds of futures. In particular, the initial co-creation of the Overton Window of current SETI research ideas by the participants and its active review by them in real time comprises the closure of the first learning loop: participants are envisioned to learn from each other about different methods and tactics that can be used to address the problem of searching for sapient extraterrestrials, by studying this instantiation of the Overton Window as a representation of the set of alternative options that are currently within the boundaries of acceptability, while also becoming aware of other options that are currently demarcated as being “radical” or “unthinkable”.

The double-loop learning effect is envisioned to occur recursively on top of the single-loop learning effect as follows. During their mental time-travels to their most desirable and their most probable futures of a SETI success, participants gradually explicate the non-fictional narratives that underlie their travels towards and beyond this purposely undefined “success”. These narratives can be expected to be related to ideas captured in the initial Overton Window. They start from a time after the dust of a SETI success has settled, from which they move backwards in imaginary time, firstly by attempting to capture the deeper significance of that SETI success and afterwards by attempting to reconstruct the timeline that led from a fictitious initial putative detection in 2026 to that SETI success. In these two mental time-travels, the type of sensemaking that is expected to be employed by participants during the activities through which they piece together the non-fictional narratives connecting a fictitious SETI success with its past and its future is detached-deliberate sensemaking. This is because these activities are envisioned to place participants into an abstract mode of engagement with the conduct of SETI to allow them to deliberately view, analyze, and interpret the conduct of SETI from the “outside”. This sensemaking type is primarily cognitive and is conducted by means of the participants’ reflections on the conceptually-represented past, present, and future regions of the temporal continuum which underlies this persistently disruptive experience of SETI success that they investigate panoramically \citep{sats}. On the contrary, during their third and final mental time-travel to the strange but not impossible SETI-relevant future of the Apparition, participants are suddenly placed in a surprising, anomalous situation, where the conduct of SETI is challenged by a phenomenon that resists interpretation and understanding and even appears to invert the expected directionality of a discovery by seemingly making the first move towards the humans’ terrain. The experiencing of this possibility demonstrates a disturbance to the supposed fixedness of the boundaries of acceptability of the initial Overton Window, as it fundamentally flips all the non-fictional narratives related to those ideas captured in that Overton Window which tacitly assume that humanity is the “discoverer” in a SETI discovery. In this third mental time-travel, the type of sensemaking that is expected to be employed by participants during the activities through which they navigate this highly fluid and troublesome SETI-relevant situation that is presently unfolding is involved-deliberate sensemaking. This is because these activities are envisioned to place participants into a contextualized mode of engagement with the conduct of SETI to allow them to deliberately get a grasp of what it is like to conduct SETI in these ambiguous circumstances from the “inside”. This sensemaking type is partly sensorial and partly cognitive and is conducted by means of the participants’ reflections on the spontaneously unfolding temporal spiral that instantly weaves together pasts and futures into the present of the immediate interruption posed to the routine conduct of SETI by the Apparition, which they experience through active probing \citep{sats}. The juxtaposition of the involved subtype of deliberate sensemaking with the detached subtype, and the recognition of how the former complements the latter in real-time evolving situations, are then expected to question the presumed rigidity of one’s construal of the conduct of SETI as a strictly-delineated activity, leading to a reconsideration of the initial co-created Overton Window of current SETI research ideas. This comprises the closure of the second learning loop: participants are envisioned to learn from each other about different fundamental assumptions and logics that can equally plausibly underlie one’s understanding of the problem of searching for sapient extraterrestrials, by studying how new and former SETI research ideas collectively move inwards or outwards the boundaries of acceptability of the new Overton Window, as compared to its initial instantiation.

Lastly, the triple-loop learning effect is envisioned to occur recursively on top of the double- and single-loop learning effects as follows. At the end of each of their mental time-travels to the three different kinds of futures, i.e., the desirable, the probable, and the strange, participants contemplate the overall “foundations” that held together their sensemaking undertakings in each future, which correspond to the clusters of anticipatory assumptions of the Futures Literacy Framework \citep{mila}. This contemplation vis-à-vis each of these three different kinds of futures in isolation from the other two is envisioned to help them learn how each anticipatory assumption cluster, or “foundation”, gets summoned to mediate access to specific kinds of futures. The latter is expected to spotlight the fact that specific motives for consciously engaging with the future, i.e., to prepare or plan or to navigate emergent novelty, can tacitly increase the salience of some particular anticipatory assumption clusters and decrease the salience of others, thereby accordingly influencing one’s ways of understanding and acting. By first distinguishing among these three different kinds of futures and contemplating each of them in isolation and then contemplating across all three in the very last activity of the workshop, participants are envisioned to learn that different kinds of futures might become tacitly intermingled in one’s future thinking, acting as inadvertently mixed motives that can unwittingly modulate one’s sensemaking, unless they are identified. It is through this last overall contemplation that participants are envisioned to learn that they can bypass the limits which those futures impose like active filters on any instantiation of the SETI field’s Overton Window, if they consciously employ active awareness to discern which kinds of futures might be jointly in force and then to purposely distinguish them in order to work with them in isolation in their own terms. In other words, this contemplation is envisioned to help participants learn the notion of “walking on two legs”, i.e., appropriately shifting between “anticipation-for-the-future” and “anticipation-for-emergence” \citep{mila}, a metaphor that captures the essence of exercising futures literacy. By increasing their futures literacy, participants are envisioned to become more adept at consciously alternating between deploying the “detached” and “involved” subtypes of deliberate sensemaking to fit the particular demands of a situation, a process that can change which entire sets of alternative affordances regarding the conduct of SETI can be perceived in the present at a given time. This comprises the closure of the third loop.

All in all, after experiencing this workshop, participants are envisioned to have learned that the range of SETI research ideas can be tacitly limited by the frames of the acceptable narratives in which they can get bundled. When left to their own devices, those research ideas can gradually bend towards less diversity of imaginings and less discernibility of factuality from desirability and wishfulness, due to the influence of SETI’s peculiar social imaginary to which their underlying narratives are ultimately connected. Therefore, a conscious engagement of SETI scholars with all those elements is required. This can enable the active shifting of the filters that color what the present conditions at any given time can afford to the conduct of SETI, not only in terms of broadening and diversifying the narratives which are getting operationalized, but also in terms of being ready to engage with ambiguous, borderline cases of strange but corroborated detections which may not clearly abide by such narratives whose structures by default encapsulate complete discoveries.

\section*{Acknowledgments}

\noindent The author would like to thank the Chair and the Members of the IAUS 404 Organising Committee for their pivotal support in evaluating the proposed workflow that was implemented in the Symposium's “Futures Workshop” session as a pilot. The author gratefully acknowledges support from the SETI Institute's Discovery and Futures Lab. Any opinions, findings, and conclusions or recommendations expressed in this material are those of the author.


\begin{thebibliography}{}
\bibitem[Benedek, 2018]{benedek}
{Benedek}, M. 2018,
\newblock {Internally Directed Attention in Creative Cognition}. \newblock In {Jung}, R.~E. \& {Vartanian}, O., editors, {\em Cambridge handbook of the neuroscience of creativity}, pp. 180--194. Cambridge University Press.

\bibitem[Castelfranchi and Miceli, 2011]{casmic2011}
{Castelfranchi}, C. \& {Miceli}, M. 2011,
\newblock {Anticipation and Emotion}.
\newblock In {Petta}, P., {Pelachaud}, C., \& {Cowie}, R., editors, {\em Emotion-oriented systems: the HUMAINE handbook}, pp. 479--496. Springer-Verlag.

\bibitem[Charbonneau, 2024]{charb}
Charbonneau, R. 2024,
\newblock SETI, artificial intelligence, and existential projection. {\em Physics Today}, 77(2), pp. 36--42.

\bibitem[Crane et al., 2010]{crane}
Crane, C., Barnhofer, T., Hargus, E., Amarasinghe, M., \& Winder, R. 2010,
\newblock The relationship between dispositional mindfulness and conditional goal setting in depressed patients. {\em British Journal of Clinical Psychology}, 49(3), pp. 281--290.

\bibitem[Dator, 2009]{dator}
Dator, J. 2009,
\newblock Alternative Futures at the Manoa School. {\em Journal of Futures Studies}, 14(2), pp. 1--18.

\bibitem[de Bono, 2000]{debono}
de Bono, E. 2000,
\newblock {\em Six Thinking Hats}.
\newblock Penguin.

\bibitem[Denning, 2017]{denn}
Denning, K. 2017,
\newblock Hawking Says that Discovering Intelligent Life Elsewhere Would Spark Greater Compassion and Humility among Us… But Why Wait?. {\em Theology and Science}, 15(2), pp. 142--146.

\bibitem[Dick, 2018a]{dicka}
{Dick}, S.~J. 2018a,
\newblock {Discovery}. 
\newblock In {\em Astrobiology, discovery, and societal impact}, pp. 37--64. Cambridge University Press.

\bibitem[Dick, 2018b]{dickb}
{Dick}, S.~J. 2018b,
\newblock {Is Human Knowledge Universal?}. 
\newblock In {\em Astrobiology, discovery, and societal impact}, pp. 141--175. Cambridge University Press.

\bibitem[French et al., 2025]{french}
French, A.~M., George, A., Madden, J. \& Storey, V.~C. 2025, \newblock Tabloids, Fake News, and the Overton Window: The COP Model on News Consumption in Uncertain Times. {\em Information Systems Frontiers}, 27, pp. 2317--2335.

\bibitem[Gardin and Roux, 2004]{garo}
Gardin, J.~-C. \& Roux, V. 2004,
\newblock The Arkeotek project: a European network of knowledge bases in archaeology of techniques. {\em Archeologia e calcolatori}, 15, pp. 25--40.

\bibitem[Hadley and MacLeod, 2010]{hama}
Hadley, S.~A. \& MacLeod, A.~K. 2010,
\newblock Conditional goal-setting, personal goals and hopelessness about the future. {\em Cognition and Emotion}, 24(7), pp. 1191--1198.

\bibitem[Haqq-Misra et al., 2020]{hmks}
Haqq-Misra, J., Kopparapu, R.~K., \& Schwieterman, E. 2020, \newblock Observational constraints on the great filter. {\em Astrobiology}, 20(5), pp. 572--579.

\bibitem[Junge, 2011]{junge}
{Junge}, K. 2011,
\newblock {Self-concepts, Counter-concepts, Asymmetrical Counter-concepts: Some Aspects of a Multi-faceted Agenda}. 
\newblock In {Junge}, K. \& Postoutenko, K., editors, {\em Asymmetrical Concepts after Reinhart Koselleck: Historical Semantics and beyond}, pp. 9--49. Transcript Verlag.

\bibitem[Kwon and Nicolaides, 2017]{kwni}
Kwon, C.~K. \& Nicolaides, A. 2017,
\newblock Managing diversity through triple-loop learning: A call for paradigm shift. {\em Human Resource Development Review}, 16(1), pp. 85--99.

\bibitem[Lemarchand, 2004]{lema}
Lemarchand, G.~A. 2004, 
\newblock The Technological Adolescent Age Transition: A Boundary to Estimate the Last Factor of the Drake Equation. {\em Symposium - International Astronomical Union}, 213, pp. 460--466.

\bibitem[Miceli and Castelfranchi, 2014]{mica}
{Miceli}, M. \& {Castelfranchi}, C. 2014,
\newblock {Anticipatory Emotions}. 
\newblock In {\em Expectancy and Emotion}, pp. 124--183. Oxford University Press

\bibitem[Miller, 2018a]{mila}
{Miller}, R. 2018a,
\newblock {Sensing and making-sense of Futures Literacy: towards a Futures Literacy Framework (FLF)}.
\newblock In {Miller}, R., editor, {\em Transforming the future: Anticipation in the 21st century}, pp. 15--50. Routledge. 

\bibitem[Miller, 2018b]{milb}
{Miller}, R. 2018b,
\newblock {Futures Literacy Laboratories (FLL) in practice: An overview of key design and implementation issues}.
\newblock In {Miller}, R., editor, {\em Transforming the future: Anticipation in the 21st century}, pp. 95--109. Routledge. 

\bibitem[Morrison, 1995]{morr}
{Morrison}, P. 1995,
\newblock {A Talk with Philip Morrison}.
\newblock In {\em Nothing is Too Wonderful to be True}, pp. 196--203. AIP Press. 

\bibitem[Moskowitz, 2002]{mosk}
Moskowitz, G.~B. 2002,
\newblock Preconscious effects of temporary goals on attention. {\em Journal of Experimental Social Psychology}, 38(4), pp. 397--404.

\bibitem[Postoutenko, 2023]{post}
{Postoutenko}, K. 2023,
\newblock {'Asymmetrical Counter-Concepts': Chances and Challenges}.
\newblock In {Postoutenko}, K., editor, {\em Beyond 'Hellenes' and 'Barbarians': Asymmetrical Concepts in European Discourse}, pp. 1--40. Berghahn Books. 

\bibitem[Profitiliotis, 2024a]{proa}
Profitiliotis, G. 2024a,
\newblock Rehearsing Post-Detection Futures: Building Futures Literacy vis-à-vis Post-Detection Considerations for a Responsible Search for Extraterrestrial Life. {\em Proceedings of the International Astronomical Union}, 20(S387), pp. 287--311.

\bibitem[Profitiliotis, 2024b]{prob}
{Profitiliotis}, G. 2024b,
\newblock {Moral Vistas to Xenic Beyonds: Fostering Moral Imagination to Pre-empt Monsterization in Future Encounters With Extraterrestrial Life}. 
\newblock In {Stelios}, S. \& {Theologou}, K., editors, {\em The Ethics Gap in the Engineering of the Future: Moral Challenges for the Technology of Tomorrow}, pp. 177--199, Emerald Publishing Limited.

\bibitem[Profitiliotis, 2025]{proc}
Profitiliotis, G. 2025,
\newblock Enrichment of the driving metanarratives shared between SETI and space sustainability through a multispecies lens. {\em Acta Astronautica}, 239, pp. 932--938.

\bibitem[Sagan, 1980]{sag}
{Sagan}, C. 1980,
\newblock {Who Speaks for Earth?}.
\newblock In {\em Cosmos}, pp. 317--346. Random House.

\bibitem[Sandberg and Tsoukas, 2020]{sats}
Sandberg, J. \& Tsoukas, H. 2020,
\newblock Sensemaking reconsidered: Towards a broader understanding through phenomenology. {\em Organization Theory}, 1, pp. 1--34.
 
\bibitem[Sani et al., 2009]{sani}
Sani, F., Herrera, M., \& Bowe, M. 2009,
\newblock Perceived collective continuity and ingroup identification as defence against death awareness. {\em Journal of Experimental Social Psychology}, 45(1), pp. 242--245.

\bibitem[Sheikh, 2020]{shei}
Sheikh, S.~Z. 2020,
\newblock Nine axes of merit for technosignature searches. {\em International Journal of Astrobiology}, 19(3), pp. 237--243.

\bibitem[Szakolczai, 2009]{szaa}
Szakolczai, A. 2009,
\newblock Liminality and experience: Structuring transitory situations and transformative events. {\em International Political Anthropology}, 2(1), pp. 141--172.

\bibitem[Szakolczai, 2017]{szab}
Szakolczai, A. 2017,
\newblock Permanent (trickster) liminality: The reasons of the heart and of the mind. {\em Theory \& Psychology}, 27(2), pp. 231--248.

\bibitem[Szanto and Montes Sánchez, 2023]{sms}
{Szanto}, T. \& {Montes Sánchez}, A. 2023,
\newblock {Imaginary Communities, Normativity and Recognition: A New Look at Social Imaginaries}. 
\newblock In {Römer}, I. \& {Stenger}, G., editors, {\em Faktum, Faktizität, Wirklichkeit: Phänomenologische Perspektiven}, pp. 197--220, Feliz Meiner.

\bibitem[Turner, 1967]{tur}
{Turner}, V.~W. 1967,
\newblock {Betwixt and Between: The Liminal Period in Rites de Passage}. 
\newblock In {\em The forest of symbols: Aspects of Ndembu ritual}, pp. 93--111, Cornell University Press.

\bibitem[van Gennep, 1960]{vang}
van Gennep, A. 1960,
\newblock {\em The Rites of Passage}.
\newblock University of Chicago Press.

\end{thebibliography}

\end{document}